# Excitonic instability of two-dimensional tilted Dirac cones

Daigo Ohki[1*], Michihiro Hirata[2*], Takehiro Tani[1], Kazushi Kanoda[3] and Akito Kobayashi[1]

[1]Department of Physics, Nagoya University, Chikusa-ku, Nagoya 462-8602, Japan.

[2]Institute for Materials Research, Tohoku University, Aoba-ku, Sendai 980-8577, Japan.

[3]Department of Applied Physics, University of Tokyo, Bunkyo-ku, Tokyo 113-8656, Japan.

* Correspondence and requests for materials should be addressed to D.O. (dohki@s.phys.nagoya-u.ac.jp) and M.H. (michihiro_hirata@imr.tohoku.ac.jp)

**The electron-electron Coulomb interaction in Dirac-Weyl semimetals harbours a novel paradigm of correlation effects that hybridizes diverse realms of solid-state physics with their relativistic counterpart. Driving spontaneous mass acquisition, the excitonic condensate of strongly-interacting massless Dirac fermions is one such example whose exact nature remains debated. Here, by focussing on the two-dimensional tilted Dirac cones in the organic salt $\alpha$-(BEDT-TTF)$_2$I$_3$, we show that the excitonic instability is controlled by a small chemical-potential shift and an in-plane magnetic field. In combined analyses based on renormalization-group approaches and ladder approximation, we demonstrate that the nuclear relaxation rate is an excellent probe of excitonic-spin fluctuations in an extended parameter region. Comparative nuclear magnetic resonance (NMR) experiments show good agreements with this result, jointly revealing the importance of intervalley nesting between field-induced, spin-split Fermi pockets of opposite charge polarities. Our work provides an accurate framework to search for excitonic instability of strongly-interacting massless fermions.**

The notable electronic properties of graphene as well as topological semimetals and insulators have been attracting increasing attention not only because of their exotic topological nature but also due to their unusual effects induced by the electron-electron Coulomb interaction[1–7]. In contrast to conventional metals, the metallic screening in these systems is absent near the Fermi energy $E_F$ when $E_F$ is tuned to the charge-neutral Dirac point owing to the vanishingly small carrier density, resulting in a preserved long-range component of the Coulomb interaction[1,2]. Remarkably, this long-range part of the interaction brings about an anomalous upward renormalization of the Fermi velocity $v$ due to a logarithmic self-energy correction akin to what has been commonly discussed in relativistic Dirac and Weyl theories[3,4]; the strength of the Coulomb interaction is characterized by a dimensionless coupling constant $\alpha = e^2/4\pi\varepsilon_0\varepsilon\hbar v$ (with the elementary charge $e$ and the relative permittivity $\varepsilon$) that mimics the fine-structure constant in quantum electrodynamics ($\alpha_{\text{QED}} \simeq 1/137$), describing motions and interactions of charged relativistic particles[8]. For a weakly-interacting regime (or weak coupling), extended studies of the self-energy effect were reported first at the level of first-order perturbative calculations[2] and later by more elaborate renormalization-group (RG) approaches[2,4,9]. For strong coupling (typically $\alpha > 1$), Monte Carlo, RG and mean-field theories in the honeycomb lattice[10–17] as well as Weyl semimetals[18–27] predicted a breakdown of the massless-fermion picture due to excitonic condensation of electron-hole pairs and a concomitant gap opening at $E_F$[3,4,28–30]. Being a cousin of the chiral symmetry breaking in high-energy physics[31], lifting the degeneracy of a pseudospin-1/2 degree of freedom projected onto the momentum (i.e., chirality)[4], this novel type of excitonic instability has been intensively studied from theoretical perspectives, but experimentally, not much is understood; in particular, its relationship with the experimental phase diagram remains entirely unknown since $\alpha$ is small in many existing systems[3,4].

In monolayer graphene having two-dimensional (2D) vertical Dirac cones, the largest ever observation of the velocity renormalization (also known as the running coupling constant) comes from a suspended device where quantum oscillation experiments revealed a sharp increase of $v$ by more than a factor of 2 upon decreasing carrier density[32]. Several other measurements are also believed to capture this renormalization using various quantities as a control parameter such as substrate's dielectric constant, magnetic field $H$ and temperature $T$. Despite extensive efforts to increase $\alpha$, all the existing studies point to a weak coupling (i.e., small $\alpha$ ($\lesssim 1$)) in good agreement with the absence of phase transitions in monolayer specimens.

By contrast, recent studies in the pressurized organic salt $\alpha$-(BEDT-TTF)$_2$I$_3$ have highlighted another example of the velocity renormalization where 2D tilted Dirac cones (Fig. 1a) are confirmed by magnetotransport[33–35], calorimetric[36] and nuclear magnetic resonance (NMR)[37,38] measurements. Among these $^{13}$C-NMR experiments in conjunction with RG analyses found a logarithmic suppression of the Knight shift $K$ upon cooling as expected for a $T$-dependent renormalization of $v$[39], signalling that the tilted Dirac cones are nonuniformly reshaped by the self-energy effect due to the unscreened long-range Coulomb interaction (Fig. 1c)[37]. Furthermore, the resultant fits to the Knight-shift data revealed a sizeable bare coupling constant of $\alpha = 12.6$, indicating that the system lies close to the strong-coupling regime; indeed, the nuclear spin-lattice relaxation rate $1/T_1$ observed an anomalous upturn at low $T$[38] that is shown by numerical simulations to represent nonzero-momentum spin fluctuations due to intervalley excitonic pairing instability, developing as a precursor to the condensation with a mass gap opening. These findings offer an excellent testing ground for studying the strong-coupling physics of interacting massless Dirac fermions in this material, both from theoretical and experimental perspectives.

Here, we theoretically investigate the excitonic phase diagram of interacting 2D massless Dirac fermions in $\alpha$-(BEDT-TTF)$_2$I$_3$, by incorporating the impacts of the spin-degeneracy splitting and a shift of the chemical potential $\mu$ off the charge-neutral Dirac points. By solving a gap equation considering ladder-type diagrams and the velocity renormalization due to the RG correction, we show that off-neutrality drastically suppresses the intervalley excitonic instability, whereas it is enhanced by an in-plane $H$ that drives a field-induced Fermi-surface nesting in the intervalley process, connecting the Fermi pockets of opposite spin and charge polarities. Furthermore, we numerically evaluate the $\mu$ and $H$ dependence of $1/T_1$ and compare it with the $^{13}$C-NMR data in $\alpha$-(BEDT-TTF)$_2$I$_3$, which provides rational ways to interpret the data in the frame of the intervalley excitonic instability in the $T$-$\mu$-$H$ phase space.

## Results

**Microscopic Model.** The 2D massless Dirac fermions generally possess two spin-degenerate Dirac points in the first Brillouin zone, protected by space and time inversion symmetry[40,41]. For the honeycomb lattice of graphene, these points are fixed at the Fermi energy $E_F$ and locate at the corners of the hexagonal zone boundary where vertical Dirac cones reside. In $\alpha$-(BEDT-TTF)$_2$I$_3$, by contrast, anisotropic networks of the hopping amplitudes[41–46] shift the charge-neutral Dirac points from high symmetry positions to general momentum ($\pm \mathbf{k}_0$), causing a tilt of the cones in opposite directions (Fig. 1a). In the presence of an in-plane $H$, the spin- degeneracy splitting generates elliptic Fermi surfaces, showing little overlap between the spin-↑ electron and spin-↓ hole pockets because of the tilt (Fig. 1b).

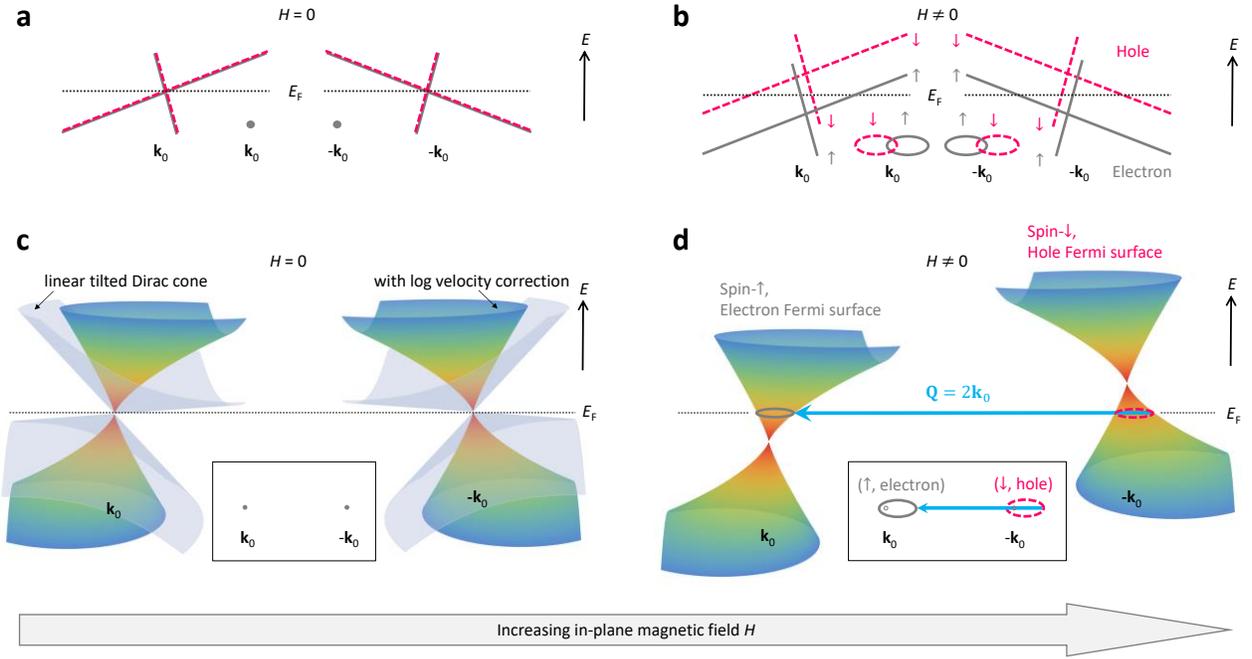

**Fig. 1** Schematic illustrations of the spin splitting and the interband Fermi-surface nesting in the intervalley process for charge-neutral 2D tilted Dirac cones. **a** Spin-degenerate titled Dirac cones at charge neutrality. A pair of 2D Dirac points anchored to the Fermi energy $E_F$ resides at general positions in the first Brillouin zone ($\mathbf{k}_0$ and $-\mathbf{k}_0$). **b** An in-plane magnetic field $H$ lifts the spin degeneracy and yields spin-↑ electron and spin-↓ hole pockets, which are elliptic and poorly overlap each other due to the opposing directions of the cone's tilt. **c** Calculated images of the non-interacting spin-degenerate cones in $\alpha$-(BEDT-TTF)$_2$I$_3$ at charge neutrality (outer cones) and the reshaped cones due to the logarithmic self-energy correction by the unscreened Coulomb interaction (inner cones), using the bare Coulomb coupling of $\alpha = 12.6$ in line with Ref. 38 (see Methods). **d** Spin-split reshaped cones due to the electron Zeeman effect in an in-plane $H$. The spin-↑ cone (left; $\mathbf{k}_0$) and the spin-↓ cone (right; $-\mathbf{k}_0$) are specifically presented. Inset of **c** and **d**: Point-like Fermi surfaces at zero field (**c**) and field-induced spin-↑ electron and spin-↓ hole pockets (**d**). The perfect interband Fermi surface nesting in the intervalley process with a momentum transfer of $\mathbf{Q} = 2\mathbf{k}_0$ (arrow) drives the intervalley excitonic instability in the even-parity, spin-transverse channel (see the text).

We consider a low-energy theory derived from an effective tight-binding model in Ref. 46 describing the 2D titled Dirac cones in $\alpha$-(BEDT-TTF)$_2$I$_3$[47–49], using so-called Luttinger-Kohn (LK) bases[50] (for details, see Methods and Ref. 37). In pristine condition, $E_F$ is anchored to the band-crossing Dirac point due to the 3/4-filling of the electronic band, but we introduce $\mu$ in our model to deal with the influence of a finite carrier-doping effect. Taking the unscreened Coulomb interaction and an in-plane $H$ into account, the effective 8×8 Hamiltonian[37,38] yields

$$H_{\text{eff}} = \sum_{\mathbf{k}} \Psi_{\mathbf{k}}^{\dagger} H_0 \Psi_{\mathbf{k}} + \sum_{\mathbf{q}} V_0(\mathbf{q}) \rho(\mathbf{q}) \rho(-\mathbf{q}), \qquad (1)$$

where $\Psi_{\mathbf{k}}^{\dagger} = \left(c_{\mathbf{k},s,\eta}^{\nu}\right)^{\dagger}$ is an 8-component creation operator describing 2D massless Dirac fermions, $\rho(\mathbf{q}) = \sum_{\mathbf{k}} \sum_{\nu,s,\eta} c_{\mathbf{k},s,\eta}^{\nu\dagger} c_{\mathbf{k}+\mathbf{q},s,\eta}^{\nu}$ is the density operator and $V_0(\mathbf{q}) = 2\pi e^2/\varepsilon|\mathbf{q}|$ is the Fourier transform of the Coulomb potential with the 2D wavenumber vectors $\mathbf{k}$ and $\mathbf{q}$. Here, the superscript, the second and third indices of the subscript represent projections of two LK pseudospin-1/2 ($\nu$ = a, b), two real spin-1/2 [$s$ = 1(↑), -1(↓)] and two valleys [$\eta$ = 1(R), -1(L)], respectively. The general form for the non-interacting Hamiltonian is given as

$$H_0 = \hbar\left(\mathbf{w} \cdot \mathbf{k} \hat{\sigma}_0 \otimes \hat{\tau}_z + v_x k_x \hat{\sigma}_x \otimes \hat{\tau}_z + v_y k_y \hat{\sigma}_y \otimes \hat{\tau}_0\right) \otimes \hat{s}_0 - \mu - \frac{g}{2} \mu_B H \hat{\sigma}_0 \otimes \hat{\tau}_0 \otimes \hat{s}_z, \qquad (2)$$

where $\mathbf{w} = (w_x, w_y)$ and $\mathbf{v} = (v_x, v_y)$ are the velocities describing the tilt and the anisotropy of the Dirac cone, respectively; $g$ = 2 is the electron g-factor in $\alpha$-(BEDT-TTF)$_2$I$_3$[51]; $\hat{\sigma}_i$, $\hat{\tau}_i$ and $\hat{s}_i$ are the Pauli matrices representing LK pseudospin-1/2, valley pseudospin-1/2 and real electron spin-1/2, respectively, with the three indices taking one of the four possible values ($i, j$ = 0, x, y and *z*). The index 0 represents a unit matrix, and we set the origin of energy at the charge-neutral Dirac point. Note that we define **k** = ($k_x$, $k_y$) only around each of the Dirac points at $\pm \mathbf{k}_0$, and omit backscattering and Umklapp processes such that both **k** and **k** + **q** are restricted around these Dirac points (i.e., **q** is much smaller compared to $2\mathbf{k}_0$).

The self-energy-induced velocity renormalization by the long-range part of the Coulomb interaction is considered within a frame of one-loop level RG calculations in the leading order in 1/*N* (*N* >>1), which is valid for both weak and strong Coulomb interaction, with *N* = 4 being the number of fermion species (i.e., two spin projections and two valleys)[2,4,9]. To be compatible with experimental $^{13}$C-NMR data[37,38], we employ the upward RG flow of the velocity **v** by a *T*-driven running coupling constant (where **w** does not flow at one-loop level)[1,2,37,39]. The dimensionless Coulomb coupling – the ratio of the Coulomb potential to the electron kinetic energy – is defined as $\alpha = e^2/\left(4\pi\varepsilon_0\varepsilon\hbar\sqrt{v_x^2 \sin^2\varphi + v_y^2 \cos^2\varphi}\right)$ for anisotropic 2D massless Dirac fermions[37], where $\varphi$ is an angle measured around the Dirac point. In *α*-(BEDT-TTF)$_2$I$_3$ the bare Coulomb coupling is approximated as $\alpha \approx e^2/4\pi\varepsilon_0\varepsilon\hbar v_0$ since the anisotropy is negligibly small ($v_x \approx v_y \equiv v_0$). Using the effective tight-binding model in Ref. 46 as our minimal starting point, RG fits to the *T*-dependent Knight shift[37] yield $\alpha$ = 12.6 at the momentum cutoff $\Lambda$ = 0.667 Å$^{-1}$ of the size of the inverse lattice constant[52] (for details, see Methods). By decreasing the energy scale from the high-energy cutoff to the low-energy region near the Dirac point (corresponding to reducing *T* in our case), $\alpha$ flows to a smaller value due to the upward velocity renormalization. At low *T* an effective value of $\alpha_{\text{eff}} \approx 1$ is necessitated for reproducing the observed upturn of $1/T_1T$ in *α*-(BEDT-TTF)$_2$I$_3$, which is relevant to the growing intervalley excitonic spin fluctuations[38].

**Intervalley excitonic instability.** For the excitonic instability in an in-plane *H*, eight types of order parameters can be considered[12,38], categorized based on the parity, spin and valley. In the charge-neutral case of 2D tilted Dirac cones with the RG correction effect considered, the excitonic order parameter with the even-parity, spin-transverse pairing in the intervalley process proves to be significant at a mean-field level treatment of gap equations[38]. This is linked to the interband Fermi surface nesting for the field-induced spin-↑ electron and spin-↓ hole pockets at different cones that strongly promotes excitonic instability by a perfect intervalley nesting with $\mathbf{Q} = 2\mathbf{k}_0$ (Fig. 1d), where $\mathbf{Q}$ is a wavenumber vector. The corresponding order parameter is given by

$$\Phi^{2\mathbf{k}_0,e}(\mathbf{k};\mathbf{q}) = \Psi^\dagger_{\mathbf{k+q}}(\hat{\sigma}_x \otimes \hat{\tau}_x \otimes \hat{s}_t)\Psi_{\mathbf{k}}$$
$$= \left(c^{a\dagger}_{\mathbf{k+q},s,R}c^{b}_{\mathbf{k},s',L} + c^{b\dagger}_{\mathbf{k+q},s,R}c^{a}_{\mathbf{k},s',L} + c^{b\dagger}_{\mathbf{k+q},s,L}c^{a}_{\mathbf{k},s',R} + c^{a\dagger}_{\mathbf{k+q},s,L}c^{b}_{\mathbf{k},s',R}\right)[\hat{s}_t]_{ss'}. \quad (3)$$

In $\alpha$-(BEDT-TTF)$_2$I$_3$ this order parameter represents some sorts of density waves, likely a mixture of charge-density wave and bond-order wave, since the LK pseudospin $\hat{\sigma}_i$ does not correspond to the real-space lattice unlike the sublattice pseudospin in graphene, but is related to a superposition of molecular orbitals associated with the nonequivalent sites in the unit cell (see Methods).

Incorporating the experimentally-determined running coupling constant[37] and assuming the effective coupling of $\alpha_{\text{eff}} = 1$ at low energy in accord with Ref. 38, we consider the dynamic spin susceptibility based on the ladder approximation, and evaluate the excitonic spin fluctuations. The dominant contribution comes from the spin-transvers fluctuations in the intervalley process ($\mathbf{Q} = 2\mathbf{k}_0$) corresponding to the order parameter (3), which is expressed by

$$\chi^{R,L}_\perp(\mathbf{q}, i\omega_m) = \sum_{\mathbf{k},\nu}[\mathcal{M}^{R,L}_{\nu\nu\bar{\nu}\bar{\nu}}\{\Lambda^\nu_+(\mathbf{k};\mathbf{q}, i\omega_m)\chi^\nu_+(\mathbf{k};\mathbf{q}, i\omega_m) + \Lambda^\nu_-(\mathbf{k};\mathbf{q}, i\omega_m)\chi^\nu_-(\mathbf{k};\mathbf{q}, i\omega_m)\}$$
$$+ \mathcal{M}^{R,L}_{\nu\bar{\nu}\bar{\nu}\nu}\{\Lambda^\nu_-(\mathbf{k};\mathbf{q}, i\omega_m)\chi^\nu_+(\mathbf{k};\mathbf{q}, i\omega_m) + \Lambda^\nu_+(\mathbf{k};\mathbf{q}, i\omega_m)\chi^\nu_-(\mathbf{k};\mathbf{q}, i\omega_m)\}]. \quad (4)$$

Here,

$$\chi^\nu_+(\mathbf{k};\mathbf{q}, i\omega_m) = -T\sum_n G^{\nu\nu}_{\uparrow,R}(\mathbf{k}+\mathbf{q}, i\varepsilon_n + i\omega_m)G^{\overline{\nu\nu}}_{\downarrow,L}(\mathbf{k}, i\varepsilon_n), \quad (5)$$

$$\chi^\nu_-(\mathbf{k};\mathbf{q}, i\omega_m) = -T\sum_n G^{\nu\bar{\nu}}_{\uparrow,R}(\mathbf{k}+\mathbf{q}, i\varepsilon_n + i\omega_m)G^{\nu\bar{\nu}}_{\downarrow,L}(\mathbf{k}, i\varepsilon_n) \quad (6)$$

are defined by the massless-Dirac-fermion Green function $\hat{G}_{s,\eta} = [G^{\nu\nu'}_{s,\eta}]$ taking into account the RG correction on the Fermi velocity (i.e., the cone reshaping in Fig. 1)[37]; the pseudospin index $\bar{\nu} = b(a)$ stands for $\nu = a(b)$; and $\varepsilon_n$ ($\omega_m$) is the fermionic (bosonic) Matsubara frequency.

Note that the form factor $\mathcal{M}^{\eta,\eta'}_{\nu_1\nu_2\nu_3\nu_4}$ in Eq. (4) is a constant complex number that is linked to a unitary transformation between the LK basis and the Fourier transform of the molecular orbitals for the nonequivalent sites in the unit cell[38] (see Methods for details). The ladder vertex $\Lambda^\nu_\pm(\mathbf{k};\mathbf{q},i\omega_m)$ is given by

$$\Lambda^\nu_\pm(\mathbf{k};\mathbf{q},i\omega_m) = 1 + \sum_{\mathbf{k}'} V_0(\mathbf{k}-\mathbf{k}')[\Lambda^\nu_\pm(\mathbf{k}';\mathbf{q},i\omega_m)\chi^\nu_+(\mathbf{k}';\mathbf{q},i\omega_m)$$
$$+\Lambda^\nu_\mp(\mathbf{k}';\mathbf{q},i\omega_m)\chi^\nu_-(\mathbf{k}';\mathbf{q},i\omega_m)]. \quad (7)$$

By separating the $\mathbf{k}$ dependence from the dependence on $\mathbf{q}$ and $\omega_m$, we have $\Lambda^\nu_\pm(\mathbf{k};\mathbf{q},i\omega_m) \simeq \Delta_\mathbf{k}\Lambda^\nu_\pm(\mathbf{q},i\omega_m) \simeq (1-\lambda)^{-1}$ at the level of random phase approximation, where $\Delta_\mathbf{k}$ and $\lambda$ are given by a linearized gap equation:

$$\lambda\Delta_\mathbf{k} = 2\sum_{\mathbf{k}'\nu} V_0(\mathbf{k}-\mathbf{k}')\Delta_{\mathbf{k}'}[\chi^\nu_+(\mathbf{k}';\mathbf{0},0)+\chi^\nu_-(\mathbf{k}';\mathbf{0},0)]. \quad (8)$$

Eq. (8) is a self-consistent equation describing the intervalley excitonic instability that favours an opening of the gap $\Delta_\mathbf{k}$ at $E_F$ as the eigenvalue $\lambda$ reaches unity.

**Phase diagram.** Based on the aforementioned formalism, one can examine the intervalley excitonic instability for given values of $\mu$ and $H$ above the transition temperature $T_c$, deduced from the $T$ point where $\lambda = 1$ is fulfilled in Eq. (8). Figure 2a, b show the calculated $T$ dependence of $\lambda$ for the even-parity, spin-transverse excitonic instability in the intervalley process ($\mathbf{Q} = 2\mathbf{k}_0$), depicted at selected values of in-plane $H$. In the charge-neutral case ($\mu = 0$) the applied $H$ remarkably enhances $\lambda$ with decreasing $T$, and yields $\lambda > 1$ at low $T$ for large $H$ (Fig. 2a). The low-$T$ rise of $\lambda$ with increasing $H$ clearly suggests an enhancement of the intervalley excitonic instability, which becomes evident by looking closer at the condition of the Fermi surface nesting. We recall that the field-induced spin-↑ electron and spin-↓ hole Fermi pockets have perfect interband nesting for $\mu = 0$ in the intervalley process (Fig. 2c). Since the applied $H$ does not alter the electron-hole symmetry, an increase in $H$ results in a widening of the sizes of the Fermi pockets but keeps the perfect nesting intact, leading to a gain in the condensation energy and helping stabilize the intervalley excitonic phase.

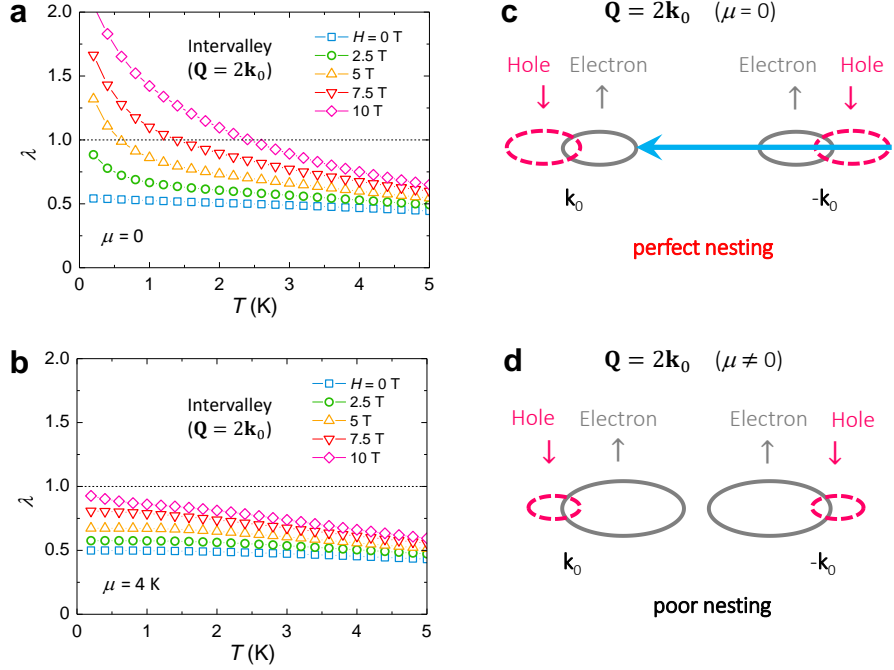

**Fig. 2** Eigen value for the intervalley excitonic instability in the 2D tilted Dirac cones of $\alpha$-(BEDT-TTF)$_2$I$_3$. **a, b** Temperature dependence of the calculated eigenvalue $\lambda$ for the even-parity, spin-transverse excitonic instability in the intervalley pairing ($\mathbf{Q} = 2\mathbf{k}_0$). The self-energy effect by the long-range part of the Coulomb interaction is considered as in Refs. 37,38. Data for a range of in-plane magnetic field $H$ are presented in the (**a**) charge-neutral case ($\mu = 0$) and the (**b**) off-neutral case ($\mu = 4$ K). **c, d** Schematic illustrations of the field-induced spin-↑ electron (solid) and spin-↓ hole (dashed) Fermi pockets for different sizes of $\mu$. The Fermi pockets have perfect interband nesting (arrow) in the intervalley process for $\mu = 0$ (with $\mathbf{Q} = 2\mathbf{k}_0$) (**c**), whereas the nesting is poor for $\mu \neq 0$ due to the large electron-hole asymmetry (**d**).

On the contrary, the increase of $\lambda$ becomes smaller at low $T$ In the charge off-neutral case ($\mu \neq 0$), retaining its size below unity even at high $H$ (Fig. 2b). This apparent suppression of $\lambda$ can be interpreted by a degraded Fermi surface nesting due to nonzero $\mu$; a finite shift of $\mu$ off the charge-neutral Dirac point causes an asymmetric size change in the field-induced Fermi pockets and substantially worsen the intervalley nesting condition (Fig. 2d), prohibiting the condensation in this process.

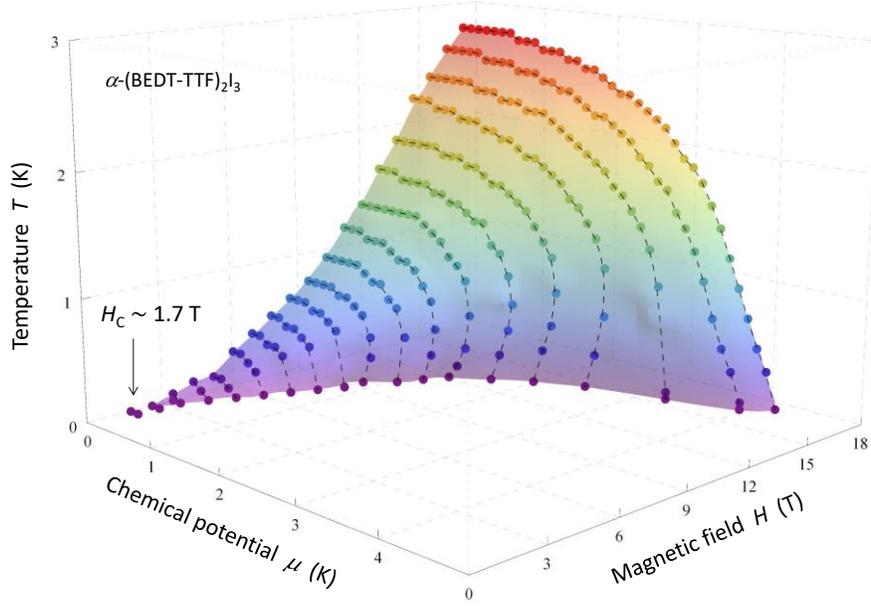

**Fig. 3** Excitonic phase diagram in $\alpha$-(BEDT-TTF)$_2$I$_3$. The calculated transition temperature $T_c$ for the interband excitonic condensation in the intervalley ($\mathbf{Q} = 2\mathbf{k}_0$) process for even-parity, spin-transverse pairing, plotted as a function of chemical potential $\mu$, in-plane magnetic field $H$ and temperature $T$. The RG correction of the velocity by the long-range part of the Coulomb interaction[37,38] is taken into account. The arrow indicates the critical field $H_C$.

To systematically evaluate the impacts of $\mu$ and $H$ on the excitonic instability, we further determine the phase diagram from the parameter dependence of $T_c$ for the even-parity, spin-transverse condensate in the intervalley ($\mathbf{Q} = 2\mathbf{k}_0$) process, as shown in Fig. 3. At low $H$ above a threshold $H_c \approx 1.7$ T, the excitonic phase appears only in a limited region at low $T$ near the charge-neutral Dirac point ($\mu = 0$), reflecting the small sizes of the field-induced Fermi pockets and the consequential little gain in the condensation energy. Upon increasing $H$, the excitonic region gradually increases to broader parameter ranges such that the condensation is favoured even in the charge off-neutral region. One can ascribe this field-driven stabilization to the enlarged sizes of the Fermi pockets at large $H$ that help to gain the condensation energy.

Moreover, the extended excitonic region to a wider $\mu$ range at higher $H$ is attributed to the improved intervalley Fermi surface nesting since the relative influence of the electron-hole asymmetry becomes smaller as the pocket sizes enlarge with increasing $H$.

We should note that this instability is rather sensitive to a small amount of $\mu$ shift especially around the charge-neutral Dirac point at $\mu = 0$. For example, at $H = 5$ T the intervalley excitonic phase disappears above $\mu = 1$ K, corresponding to an electron doping of just a few ppm of the conduction band. This small $\mu$ shift goes along with its experimentally reported size in the pressurized $\alpha$-(BEDT-TTF)$_2$I$_3$ distributing in a range of a few Kelvin among different samples, as reported by Hall measurements and supported by associated model calculations[53,54]. The notable sensitivity of the excitonic state to $\mu$ provides another support to the notion that the intervalley nesting condition between the field-induced Fermi pockets crucially affects the excitonic instability.

**Nuclear spin-lattice relaxation rate.** In the previous $^{13}$C-NMR studies[37,38] experiments found, and theories confirmed, that the low-energy spin excitations in the pressurized $\alpha$-(BEDT-TTF)$_2$I$_3$ consist of two parts: The intravalley excitations around each Dirac point at $\pm\mathbf{k}_0$, probed by the uniform part ($\mathbf{Q} = 0$) of the electron spin susceptibility, and the intervalley excitations between the two Dirac points, appearing in the $\mathbf{Q} \approx 2\mathbf{k}_0$ response of the susceptibility. These excitations can be investigated by $1/T_1T$ that is proportional to the $\mathbf{Q}$ average of the imaginary part of the transverse dynamic spin susceptibility $\mathrm{Im}\chi_\perp(\mathbf{Q}, \omega)$[55] (see Methods). Remarkably, the $\mathbf{Q} = 0$ part is strongly suppressed at low $T$ by the upward renormalization of the Fermi velocity due to the long-range part of the Coulomb interaction, whereas its impact is rather limited for the $\mathbf{Q} \approx 2\mathbf{k}_0$ response[38]. Consequently, the low-$T$ relaxation rate becomes extremely sensitive to the $\mathbf{Q} \approx 2\mathbf{k}_0$ excitations, making it particularly appealing to the study of the intervalley excitonic instability. Combined with the above knowledge of the phase diagram (Fig. 3), we are thus motivated to numerically evaluate $1/T_1T$ in a wide parameter region, and check the impacts of $\mu$- and $H$-changes on the previously reported upturn of $1/T_1T$, induced by the intervalley excitonic spin fluctuations[38].

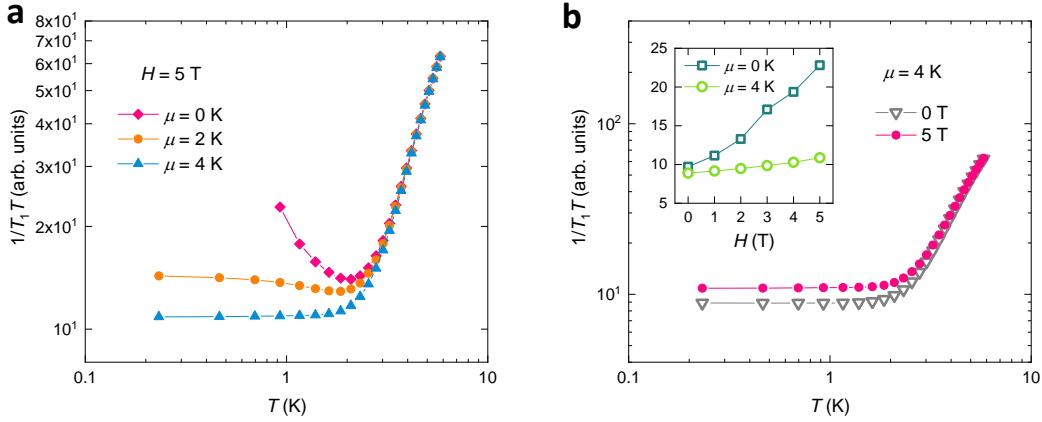

**Fig. 4** Numerical calculation of $1/T_1T$. **a** $1/T_1T$ plotted as a function of $T$ in an in-plane magnetic field of $H$ = 5 T for the chemical potential of $\mu$ = 0, 2 and 4 K. **b** Temperature dependence of $1/T_1T$ at $\mu$ = 4 K for $H$ = 0 and 5 T. Inset of **b**: Field dependence of $1/T_1T$ at $T$ = 0.2 K for $\mu$ = 0 and 4 K. The ladder vertex [Eq. (7)] and the self-energy correction by the long-range part of the Coulomb interaction[37,38] are considered.

Figure 4a, b show the calculated $T$ dependence of $1/T_1T$ for several values of $\mu$ and $H$, incorporating both the RG correction and the excitonic-pairing instability driven by the ladder vertex (Eq. (7)). In the charge-neutral case ($\mu$ = 0) with an in-plane $H$ of 5 T, $1/T_1T$ shows a clear upturn at low $T$ by the growing excitonic spin fluctuations[38] (Fig. 4a). A small increase of $\mu$ rapidly suppresses the upturn, and $1/T_1T$ at low $T$ eventually levels off for $\mu$ = 4 K. This clearly signals that the intervalley excitonic instability is weakened upon increasing $\mu$, in excellent agreement with the absence of the condensation in the large $\mu$ and small $H$ region in the phase diagram (Fig. 3). Note that this kind of low-$T$ levelling-off of $1/T_1T$ was previously reported for $\mu$ = 0 in the absence of the ladder vertex[38], but we show here, at finite $\mu$, that it becomes dominant even in its presence: We find that this phenomenon is a universal characteristic of the interband electron-hole excitations for the intervalley process in 2D Dirac cones, which dominates the $T$ dependence of $1/T_1T$ when the excitonic-pair formation is hampered by a poor Fermi surface nesting (see Supplementary Note 1 and Supplementary Fig. 1 for details).

In Fig. 4b we present $1/T_1T$ data for $\mu = 4$ K at different values of $H$. Little $H$ dependence is observed in the levelling-off-like behaviour of $1/T_1T$ in agreement with the small impact of $H$ on the $T$ dependence of $\lambda$ for $\mu \neq 0$ (Fig. 2b). This should be contrasted to the $\mu = 0$ case where $1/T_1T$ at low $T$ monotonically increases with increasing $H$ (inset of Fig. 4b) as a direct consequence of the low-$T$ enhancement of $\lambda$ at higher $H$ (Fig. 2a). The larger $H$ dependence for smaller $\mu$ is perfectly accounted for by the field-induced, intervalley Fermi surface nesting since a better nesting at higher $H$ provides stronger excitonic instability and therefore larger spin fluctuations (see Fig. 2c, d).

**Discussion**

Keeping all these insights in mind, one would immediately be tempted to make a direct comparison of theory and experiment by controlling the three parameters: $T$, $H$ and $\mu$. Varying the first two external parameters ($T$ and $H$) is experimentally easy and is fruitful to this end. But since our discussion concerns the layered charge-transfer salt $\alpha$-(BEDT-TTF)$_2$I$_3$, comprised of alternatingly-stacked 2D conducting cation (BEDT-TTF$^{+0.5}$) layers and insulating anion (I$_3^-$) layers[56], changing $\mu$ in an arbitrary and homogeneous fashion is practically not that easy. Indeed, quantum oscillation measurements in a thin crystal of $\alpha$-(BEDT-TTF)$_2$I$_3$ revealed that carrier injection is possible by electrostatic means but only into its top few layers, leaving $\mu$ unaltered in the vast majority of the crystal[57]. (Note that the system is almost purely 2D since the conductivity in the 2D plane is $10^3$ times larger than the out-of-plane conductivity[56].)

Notwithstanding the restriction, previous magnetotransport measurements in $\alpha$-(BEDT-TTF)$_2$I$_3$ have reported a small distribution of $\mu$ around the charge-neutral Dirac point, occurring randomly among different samples; in some samples, this leads at low $T$ to a sharp sign change in the Hall coefficient[54,58] and the thermopower[59,60], which is ascribed to a small electron-hole asymmetry and a sample-dependent self-doping effect of the size of $\sim$ppm of the electronic band, induced probably by tiny I$_3^-$ vacancies[53]. More recent studies suggest that moderate annealing results in an increase of the carrier density and an enhanced conductivity[58,61], which are believed to be induced by thermal desorption of triiodide I$_3$ molecules[62]. We speculate that a small deficiency in the anion layers, introduced during preparations, must be responsible for this natural self-doping, although its ppm-order size is small enough to safely forget about the associated disorder problems. Given our results in Fig. 4, the $^{13}$C-NMR relaxation rate proves to be a particularly sensitive probe of this natural $\mu$-distribution effect near the charge-neutral Dirac point. Thus, checking the sample dependence of $T_1$ will offer an indirect yet more important opportunity to investigate the influence of $\mu$ shift upon the excitonic instability.

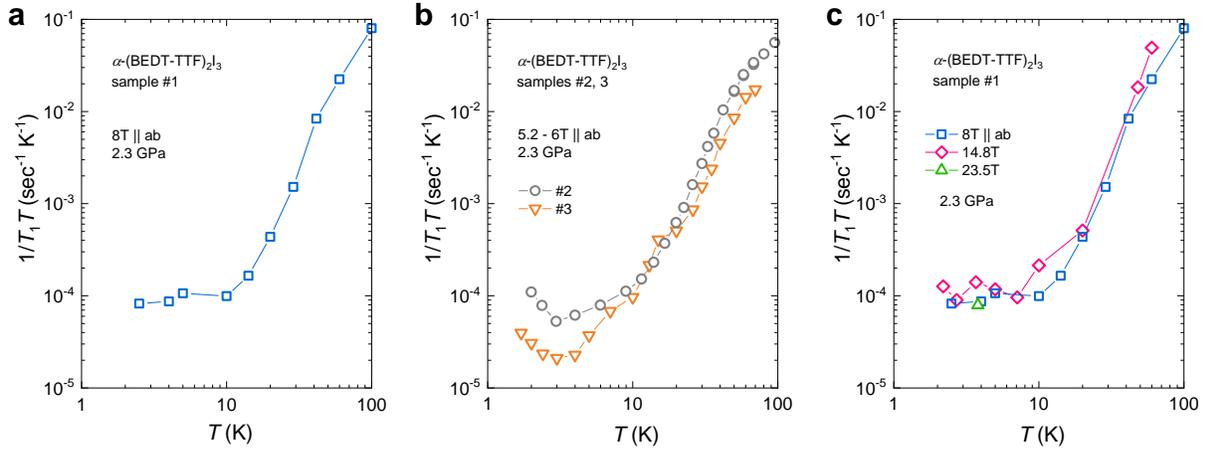

**Fig. 5** Temperature dependence of $1/T_1T$ in $\alpha$-(BEDT-TTF)$_2$I$_3$ measured by $^{13}$C NMR for three different samples (#1-3) at 2.3 GPa in an in-plane magnetic field $H$. **a**, **b**, $1/T_1T$ plotted against temperature at $H$ = 8 T in sample #1 (**a**) and 5.2 T (6T) in sample #2 (#3) (**b**). **c** Temperature dependence of $1/T_1T$ at selected values of $H$ in sample #1.

Figure 5a, b present the $T$ dependence of $1/T_1T$ measured at $^{13}$C sites in three representative $\alpha$-(BEDT-TTF)$_2$I$_3$ samples at 2.3 GPa (labelled samples #1-3), where an in-plane $H$ of 8 T (#1), 5.2 T (#2) and 6 T (#3) are applied parallel to the 2D conducting plane ($H||ab$) (see Methods). In line with our previous study in sample #3[38], $1/T_1T$ in sample #2 continues to decrease with cooling and shows an abrupt upturn below 3 K, which is ascribed to the growing intervalley excitonic spin fluctuations (Fig. 5b). In sample #1, by contrast, $1/T_1T$ shows no upturn but a levelling-off-like behaviour below 10 K (Fig. 5a), as one expects for the interband electron-hole excitations in the absence of the intervalley excitonic instability (Supplementary Figs. 1 and 2). The observation of contrasting low-$T$ behaviours in $1/T_1T$ draws an excellent qualitative

parallel with the calculation in Fig. 4a, noticeably signalling the presence of a $\mu$ shift off the Dirac point that is larger in sample #1 than in samples #2 and #3. This point is further verified by the high-field measurements in sample #1 at $H$ = 14.8 and 23.5 T (Fig. 5c) in which no appreciable $H$ dependence is observed at low $T$, in excellent agreement with the calculated $1/T_1T$ at finite $\mu$ (Fig. 4b). All our data thus provide strong support to the predicted results and coherently reinforce the idea that a shred of $\mu$ shift drastically suppresses the intervalley excitonic instability.

It should be noticed that the intravalley electron-hole pairing (with $\mathbf{Q} \approx 0$), omitted so far, barely contributes to the excitonic instability even for $\mu$ = 0 not only because of the poor $\mathbf{Q} \approx 0$ Fermi surface nesting due to the cone's tilt (see Fig. 1b)[38], but also by a suppression mechanism associated to the chirality property of the massless Dirac fermions; the electron-hole excitations, generally described by the so-called bubble-type diagrams, acquire a chirality factor near the Dirac point that scales to $C_{\delta\delta'}(\mathbf{k}; \mathbf{Q}) = [1 + \xi\delta\delta' \cos\theta_{\mathbf{k},\mathbf{k}+\mathbf{Q}}]/2$, where $\delta = \pm 1$ is the band index, $\xi = \pm 1$ refers to the intervalley or intravalley pairings and $\theta_{\mathbf{k},\mathbf{k}+\mathbf{Q}}$ is the angle between $\mathbf{k}$ and $\mathbf{k} + \mathbf{Q}$[63]. For the intravalley pairing ($\xi = 1$), this chirality factor vanishes in the interband ($\delta = -\delta'$) process with $\theta_{\mathbf{k},\mathbf{k}+\mathbf{Q}}$ = 0 that has relatively good nesting, and prevents the growth of the $\mathbf{Q} \approx 0$ excitonic instability. For the intervalley pairing ($\xi = -1$), by contrast, the corresponding interband process has perfect nesting (Fig. 2c), and the chirality factor becomes unity, which largely favours the growing excitonic instability in the $\mathbf{Q} \approx 2\mathbf{k}_0$ channel. Constructing models relying on the intervalley ($\mathbf{Q} \approx 2\mathbf{k}_0$) contribution thus provides a reliable starting point for discussing the excitonic instability of 2D tilted Dirac cones. [The chirality factor is implicitly considered in Eqs. (5) and (6).]

We comment that the previously proposed coexistence of the 2D massless Dirac fermions with ordinary massive electrons in some (but not all) $\alpha$-(BEDT-TTF)$_2$I$_3$ samples[52,64,65] does not comply with our NMR results; this picture would lead to a monotonic increase of $1/T_1T$ upon cooling due to a large saddle-point van Hove singularity near $E_F$[52] (Supplementary Note 2 and Supplementary Fig. 3) which, however, clearly opposes the observed monotonic decrease of $1/T_1T$ towards low $T$ (Fig. 5). Therefore, this comparison convincingly assures our model analyses that solely pertain to the 2D massless Dirac fermions.

Our results in Figs. 4 and 5 thus jointly indicate that the ladder approximation in our model offers a reasonable minimal framework to capture the essential nature of the intervalley excitonic instability in 2D tilted Dirac cones, at least qualitatively. To go a step ahead, one should take further care in particular of the treatments of the self-energy: The frequency dependence is one such thing that was only phenomenologically considered here [as a bandwidth reduction effect due to the on-site repulsive interaction[66], necessitated for obtaining a good quantitative estimate of $\alpha$ from the fits to the Knight shift[37] (for details, see Methods)]. The imaginary part would be another ingredient to be considered that is also omitted in the present RG model analysis. Concerning these effects probably helps to account for the discrepancy in the onset $T$ of the levelling-off between theory and experiment. Moreover, Eq. (4) is calculated within the ladder approximation in which the ladder vertex $\Lambda^v_\pm(\mathbf{k}; \mathbf{q}, i\omega_m)$ [Eq. (7)] is decomposed into the $\mathbf{k}$-dependent part ($\Delta_\mathbf{k}$) and the $\mathbf{q}$- and $\omega_m$-dependent part ($\Lambda^v_\pm(\mathbf{q}, i\omega_m)$). Using higher-level theories than this can incorporate higher order fluctuations neglected in our model, and may be valuable for further quantitative comparison of the onset $T$ in the upturn of $1/T_1T$.

The remarkable excitonic instability characterized in this study is directly linked to the chiral nature of the massless Dirac fermions [Eq. (2)] that is ubiquitous in general Dirac-Weyl systems in any dimension and for whatever types of pseudospin and symmetry[3,4,63]. Therefore, the present study offers a basic platform for qualitative understanding of the excitonic instability in widespread Dirac-cone materials, which casts new light on the interplay of the diverse physics of strongly-correlated electrons and chirality-related phenomena in various topological materials.

**Methods**

**Low-energy effective model.** We performed model calculations based on the low-energy Hamiltonian in Eq. (2), described by the so-called Luttinger-Kohn (LK) representation[50]. This representation is characterized by a unitary transformation between the LK basis, $|\psi_\mu^\eta(\mathbf{K})\rangle$, and the Fourier transform of the molecular orbital at the site $j$ in the unit cell, $|\varphi_j(\mathbf{K})\rangle$ [$j$ = A(1), A'(2), B(3) and C(4)], which is given as[42,47,48,67]

$$|\psi_\mu^\eta(\mathbf{K})\rangle = \sum_{j=1}^{4} S_{j,\mu}(\mathbf{K'}_{0,\eta})|\varphi_j(\mathbf{K})\rangle, \tag{9}$$

where $\mathbf{K}$ is an arbitrary 2D wavenumber vector defined in the first Brillouin zone, and $\mu$ (= $a, b, c$ and $d$) is an index in the LK representation at $\mathbf{K'}_{0,\eta}$. ($\mathbf{K'}_{0,\eta}$ is a fixed wavenumber defined in an infinitesimal vicinity of the right [$\eta$ = 1($R$)] or left [$\eta$ = -1($L$)] Dirac points locating at $\eta\mathbf{k}_0$). $S_{j,\mu}(\mathbf{K'}_{0,\eta})$ is described by the eigenvectors of a 4×4 tight-binding Hamiltonian, defined in a space spanned by $|\varphi_j(\mathbf{K})\rangle$ ($j$ = 1 – 4)[46,49]. For simplicity the spin index is omitted. The LK representation in Eq. (9) is smoothly connected to the ordinary Bloch representation with a unitary transformation[67]. The effective Hamiltonian given in Eq. (2) is obtained by a linearization of the 4×4 Hamiltonian around $\mathbf{K'}_{0,\eta}$ ($\eta = \pm 1$) written in the LK representation, and extracting only the $\mu = a$ and $\mu = b$ components [46]. [These two components are linked to the matrix $\hat{\sigma}_i$ in Eqs. (2) and (3) and the index $\nu$ in Eqs. (4) – (8) and (10).] Note that the form factor in Eq. (4)

$$\mathcal{M}_{\nu_1\nu_2\nu_3\nu_4}^{\eta,\eta'} = C_{\nu_1}^\eta \left(C_{\nu_2}^\eta\right)^* C_{\nu_3}^{\eta'} \left(C_{\nu_4}^{\eta'}\right)^* \tag{10}$$

is a constant complex number associated to $C_\mu^\eta \equiv S_{j=1,\mu}(\mathbf{K'}_{0,\eta})$, which is a projection of $|\psi_\mu^\eta(\mathbf{K})\rangle$ onto the $j$ = A(1) site orbital $|\varphi_{j=1}(\mathbf{K})\rangle$.

**Experimental.** Single crystals of $\alpha$-(BEDT-TTF)$_2$I$_3$[56] were synthesized from $^{13}$C-enriched BEDT-TTF molecules by conventional electrochemical methods. For $^{13}$C-NMR measurements the central carbon atoms of BEDT-TTF molecules were 99 % enriched by carbon-13 ($^{13}$C) isotopes (having a nuclear spin $I$ = 1/2). A hydrostatic pressure of $P$ = 2.3 GPa was applied to the sample with a BeCu/NiCrAl clamp-type pressure cell (C&T Factory) and the Daphne 7373 oil (Idemitsu). Samples were slightly heated up during pressurization at approximately 50 °C in order to prevent solidification of the oil[68,69] (for details, see the Methods in Refs. 37,38). A static magnetic field of 5.2 – 23.5 T was applied parallel to the crystalline *ab* plane for the $^{13}$C-NMR measurements, using standard superconducting magnets (University of Tokyo) and cryogenic-free superconducting magnets (Institute for Materials Research, Tohoku University).

$^{13}$C-NMR signals were collected using standard spin-echo techniques with commercially available homodyne spectrometers. The echo signals were recorded at a fixed radio frequency after the conventional spin-echo pulse sequence, and were converted into $^{13}$C-NMR spectra via Fourier transformation. Nuclear spin-lattice relaxation rate $1/T_1$ was determined from standard single-exponential fits to the recovery curve of nuclear magnetization after saturation, which is associated to the imaginary part of the transverse dynamic spin susceptibility [Eq. (4)] via the following expression[55]:

$$\frac{1}{T_1 T} = \frac{2\gamma_n^2 k_B (\bar{A}_\perp)^2}{g^2 \mu_B^2} \sum_{\mathbf{Q}} \frac{\mathrm{Im}\chi_\perp(\mathbf{Q},\omega)}{\omega}, \qquad (11)$$

where $\gamma_n$ is the nuclear gyromagnetic ratio, $k_B$ is the Boltzmann constant, $\mu_B$ is the Bohr magneton, $\bar{A}_\perp$ is the transverse component of the mean hyperfine coupling tensor[38] and $\omega$ is the NMR resonance frequency. Site-selective $1/T_1$ values can be determined from 8 $^{13}$C-lines, corresponding to four different molecular sites in the unit cell (*j* = A, A', B and C) split by the nuclear dipole interaction (for the details of line assignments, see Refs. 37,38). The value of $1/T_1$ is, however, site-independent below ≈ 30 K probably because of spin diffusion ensured by the nuclear spin-spin coupling among the different molecular sites. This gives rise to a unique spin temperature inside the unit cell[55] and a resultant inter-site averaging of $T_1$. Following the previous study[38], we focus on the $T_1$ data measured at the *j* = A (= A') lines as they probe the excitations around a Dirac point most uniformly[37,46,67].

**Renormalization-group analyses.** For dealing with the self-energy correction by the long-range part of the Coulomb interaction, we use the renormalization-group (RG) equations derived for the effective Hamiltonian [Eq. (2)] in the leading order in $1/N$ (with $N \gg 1$), which are valid for arbitrary strengths of the Coulomb interaction. Starting from the energy-momentum dispersion around the Dirac points obtained by diagonalization of Eq. (2)

$$E_{\pm}(q) = \hbar\left(\mathbf{w} \cdot \mathbf{q} \pm \sqrt{v_x^2 q_x^2 + v_y^2 q_y^2}\right), \qquad (12)$$

the one-loop level RG flow equations for $\mathbf{v} = (v_x, v_y)$ are given by[37]

$$\frac{1}{v_x}\frac{dv_x}{dl} = \frac{8}{\pi^2 N}\int_0^{2\pi}\frac{d\varphi}{2\pi} 2\cos^2\varphi\, F(g_\varphi),$$

$$\frac{1}{v_y}\frac{dv_y}{dl} = \frac{8}{\pi^2 N}\int_0^{2\pi}\frac{d\varphi}{2\pi} 2\sin^2\varphi\, F(g_\varphi). \qquad (13)$$

Here, $N = 4$ is the number of fermion species reflecting the two Dirac points (at $\pm \mathbf{k}_0$) and two spin projections, $\mathbf{q} = q(\cos\varphi, \sin\varphi)$ is defined around $\mathbf{k}_0$, $l = ln(\Lambda/q)$ is a momentum scale measured in the unit of the momentum cutoff $\Lambda = 0.667$ Å$^{-1}$ of the size of the inverse lattice constant[52] and circular around the Dirac point, $F(g_\varphi)$ is a function of the form $F(g_\varphi) = \left(-\pi/2 + g_\varphi + \arccos g_\varphi/\sqrt{1 - g_\varphi^2}\right)/g_\varphi$ with $g_\varphi = 2\pi e^2 N/\left(16\varepsilon\sqrt{v_x^2\sin^2\varphi + v_y^2\cos^2\varphi}\right)$ and $\varepsilon$ is the dielectric constant. ($\mathbf{w}$ does not flow at the one-loop level and stays constant.)

Using Eq. (13), fits to the $^{13}$C-NMR local-electron spin susceptibility (i.e., Knight shift) provide an excellent experimental estimate for the RG flow of the velocities[37,38]. Following Ref. 37, we use the effective tight-binding (TB) model in Ref. 46 as our minimal starting point of RG analyses, and introduce a phenomenological parameter $u$ to adjust the initial velocity values at the cutoff ($q = \Lambda$): $\mathbf{v} = u*\mathbf{v}^{TB}$, $\mathbf{w} = u*\mathbf{w}^{TB}$. Optimizing the two parameters $\varepsilon$ and $u$ by least square fits, we get $(\varepsilon, u) \approx (30, 0.35)$ (The results of the fits do not change much for $\varepsilon = 1$-30). That we have $u$ smaller than unity suggests a reduced electronic bandwidth, which is similar to what has been often discussed in correlated electron materials via dynamical mean-field methods, and ascribed to the frequency dependence of the self-energy due to the short-range part of the Coulomb interaction[66]. The bare coupling constant is then evaluated as $\alpha \approx$ $e^2/4\pi\varepsilon_0\varepsilon\hbar v_0 = 12.6$ with $v_0 = uv^{\text{TB}}$ [using $v^{\text{TB}}$ = 2.4×10$^4$ m s$^{-1}$ [46]]. For the detail of the RG flow of the coupling constant, see Supplementary Information of Ref. 38.

**Temperature dependence of the chemical potential.** To get a compatible model with transport measurements in the literature over an extended *T* range, one might need to introduce the *T* dependence of the chemical potential $\mu$ in Eq. (2), as has been suggested by Hall measurements and associated theories[53,54]. This *T* dependence appears to stem from the saddle-point van Hove singularity in the valence band, locating approximately 12 meV below the charge-neutral Dirac point as first predicted by density functional calculations[46] and later confirmed by $^{13}$C-NMR and interlayer magnetoresistance measurements[37,70]. Below 10 K, however, $\mu$ is expected to show only a minor variation upon changing *T*[53] reflecting the high electron-hole symmetry near the Dirac point. Since the *T* range of our interest lies in this low-*T* region, we can safely neglect the *T* dependence of $\mu$ in our analysis and treat $\mu$ as a phenomenological control parameter in search of excitonic instability.

**Acknowledgments**

The authors acknowledge G. Matsuno for helping with the modelling and coding; M. O. Goerbig, D. Basko, H. Yasuoka, M. Baenitz and H. Fukuyama for fruitful discussions; and M. Tamura and K. Miyagawa for providing samples and experimental supports. Part of this work was performed at the High Field Laboratory for Superconducting Materials, Institute for Materials Research, Tohoku University. This work is supported by MEXT/JSPJ KAKENHI under Grant Numbers 15K05166, 18H05225, 19J20677 and 19H01846.

# Supplementary Information

## Supplementary Figures

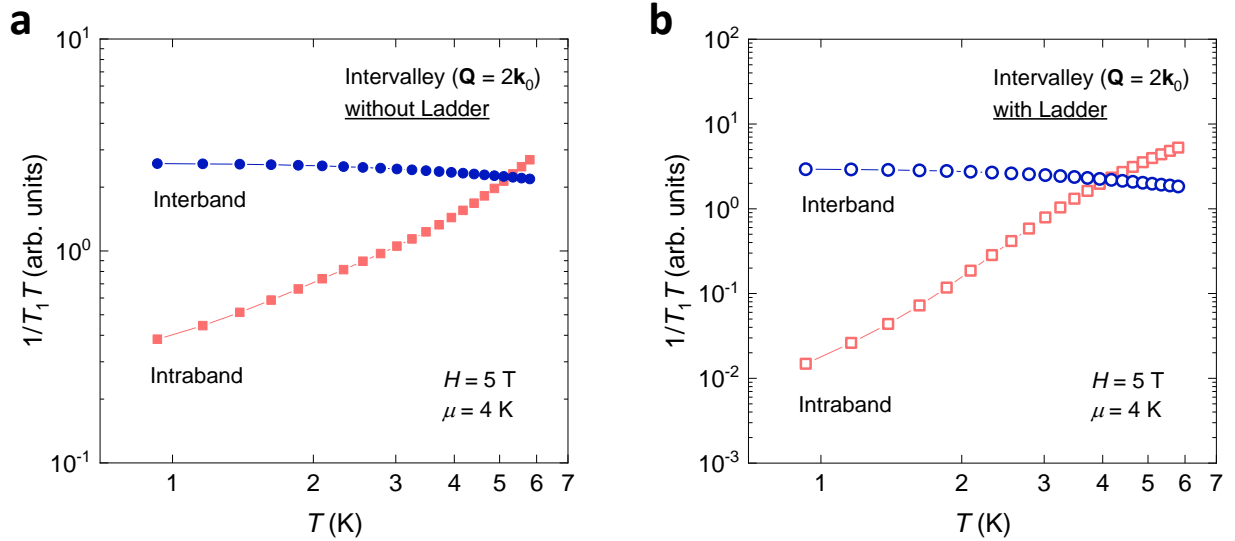

**Supplementary Fig. 1** Numerical results for the intervalley part ($\mathbf{Q} = 2\mathbf{k}_0$) of $1/T_1T$ vs $T$ for a charge off-neutral tilted Dirac cones ($\mu$ = 4 K) in an in-plane magnetic field of $H$ = 5 T. **a, b** Calculated data of the intervalley contributions of $1/T_1T$ in the absence (**a**) and presence (**b**) of the Ladder vertex [Eq. (7)] for the interband (circles) and intraband (squares) processes. The RG correction effect due to the long-range part of the Coulomb interaction is considered following Refs. 1,2 (see Methods).

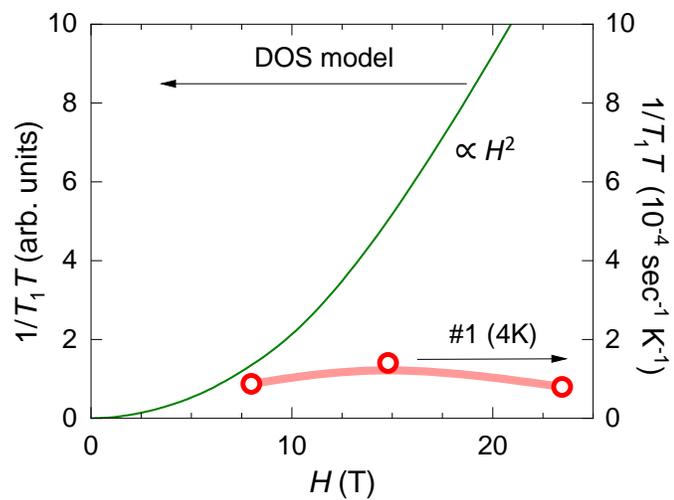

**Supplementary Fig. 2** Field dependence of $1/T_1T$ in $\alpha$-(BEDT-TTF)$_2$I$_3$. The $^{13}$C-NMR data for sample #1 at 4 K (symbols) are plotted as a function of the in-plane magnetic field $H$ together with the calculated curve following Eq. (S1) (solid line) that gives an approximate relation between $1/T_1T$ and the electron density of state (DOS). Note that these data are extracted from Fig. 5c.

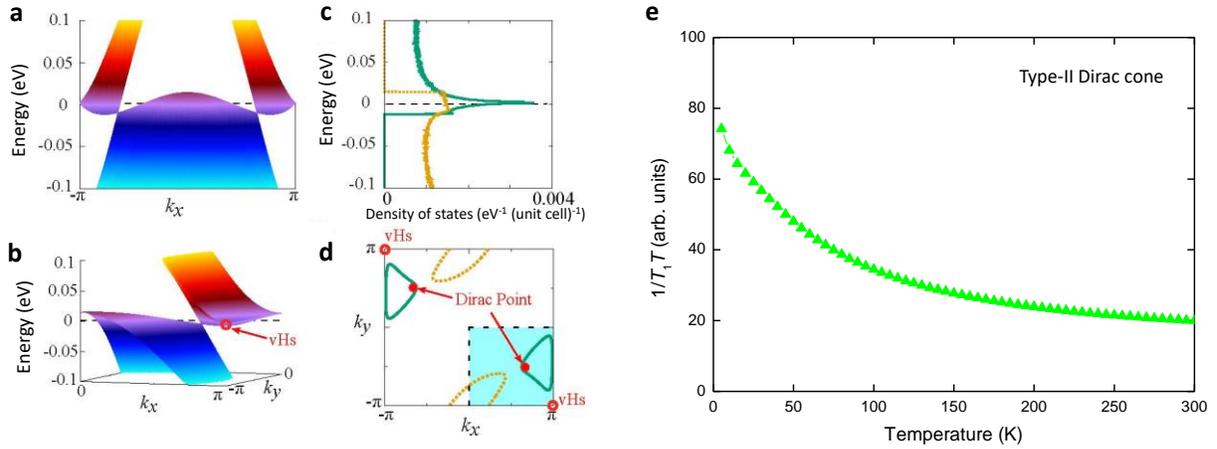

**Supplementary Fig. 3** Calculated $1/T_1T$ based on the tight-binding band structure of Ref. 3. **a** The tight-binding band structure near $E_F$ (dashed line) seen from the $k_y$ direction in $\alpha$-(BEDT-TTF)$_2$I$_3$, calculated from the X-ray diffraction data at 1.76 GPa[3]. **b** The Type-II Dirac semimetal (over-tilted cone) deduced from this model, where the Dirac point locates below $E_F$ (dashed line). At the corner $M(\pi, -\pi)$, a van Hove singularity (vHs) is present in the upper band (indicated by an arrow). **c** Density of states for the upper (green curve) and lower (orange curve) bands around $E_F$. **d** The first Brillouin zone in this model with the Fermi surfaces shown for the upper band (green closed loop) and the lower band (orange open loop). The positions of the Dirac points and vHs are indicated. The highlighted quarter of the full Brillouin zone at the right bottom (surrounded by dashed lines) corresponds to the plotted region in **b**. **e** The calculated temperature dependence of $1/T_1T$ [at the sublattice $A$ (= $A'$)] for this band structure.

## Supplementary Note 1

**Levelling-off of $1/T_1T$.** In the simplest approximation, the nuclear spin-lattice relaxation rate $1/T_1$ in conductors is expressed by means of the electron density of states (DOS) as[4,5]:

$$\frac{1}{T_1} = \frac{\pi}{\hbar}(\gamma_e\gamma_n\hbar^2)^2(A_\perp)^2 \int_{-\infty}^{\infty} dE\, D_+(E)f(E) \int_{-\infty}^{\infty} dE'\, D_-(E')\{1-f(E')\}\delta(E-E'), \quad \text{(S1)}$$

where $A_\perp$ is the hyperfine coupling constant in the direction normal to external magnetic field $H$, $\gamma_e$ ($\gamma_n$) is the electronic (nuclear) gyromagnetic ratio, $D_\pm(E) = D(E \mp E_Z/2)/2$ is the electron DOS for up/down (+/-) spins mutually shifted by the electron Zeeman energy $E_Z = g\mu_B H$, $f(E) = 1/[1+\exp\{(E-\mu)/k_B T\}]$ is the Fermi-Dirac distribution function and $\delta(E)$ is the Dirac delta function.

In the absence of the electron-electron Coulomb interaction, the linear dispersion of the 2D Dirac cones results in a linear DOS around the charge-neutrality point (set as $E = 0$), $D(E) \propto |E|$, which for small $H$ and charge-neutral case ($\mu = 0$) leads to a power law behaviour in the relaxation rate: $1/T_1 \propto T^3$.[6,7] At low $T$ the thermal energy ($k_B T$) becomes comparable to $E_Z$, and the Zeeman-induced, spin-split Fermi pockets start to play major roles, inducing a crossover upon cooling from the $T^3$ behaviour to the so-called Korringa law expected for the canonical Fermi liquid: $1/T_1 \propto T$.[6] In the latter case, simple calculations read $1/T_1 T \propto H^2$ from Eq. (S1) that is, however, in apparent opposition to the experimental data in the pressurized α-(BEDT-TTF)$_2$I$_3$ (sample #1) where little $H$ dependence is observed (Supplementary Fig. 2). This demonstrates that Eq. (S1) is invalid at low $T$ in our sample, necessitating us coming back to the general relation between $1/T_1$ and the wavenumber $\mathbf{Q}$ average of the imaginary part of the transverse spin susceptibility $\text{Im}\chi_\perp(\mathbf{Q},\omega)$ [i.e., Eq. (11)][4,5]. The breakdown of Eq. (S1) signals that the interband electron-hole excitations within each cone ($\mathbf{Q} \approx 0$) and connecting two cones at $\mathbf{k}_0$ and $-\mathbf{k}_0$ ($\mathbf{Q} \approx 2\mathbf{k}_0$; as in Fig. 1) become of paramount importance, which are omitted in Eq. (S1) but incorporated in Eq. (11).

In the previous numerical study at $\mu = 0$ considering the RG flow of the Fermi velocity in α-(BEDT-TTF)$_2$I$_3$,[1] we reported, based on Eq. (11), that the interband electron-hole excitations in the intervalley process ($\mathbf{Q} \approx 2\mathbf{k}_0$) cause little $T$ dependence in $1/T_1 T$, whereas their intraband counterparts lead to a rapid drop of $1/T_1T$ with cooling[2]. By resizing the energy and momentum scales to be compatible with the RG flow (near the Dirac point) and assuming the effective coupling of $\alpha_{\text{eff}} = 1$, we previously focussed on the low-energy regions and showed that an inclusion of the ladder vertex [Eq. (7)] brought about an upturn of $1/T_1T$ at low *T*, owing to the growing intervalley excitonic instability[2]. We extend the latter calculation to the case of $\mu \neq 0$ in this study, and find that the levelling-off-like behaviour of $1/T_1T$ universally appears due to the interband electron-hole excitations in the intervalley process ($\mathbf{Q} \approx 2\mathbf{k}_0$), regardless of the presence or absence of the ladder vertex (Supplementary Fig. 1a, b) when the excitonic instability is suppressed by a poor Fermi surface nesting (as in Fig. 2d).

## Supplementary Note 2

**Possibility of Type-II Dirac semimetals.** As we mentioned in the main text, some recent studies suggested the possibility that ordinary massive electrons might coexist with the 2D massless Dirac fermions in some (but not all) pressurized $\alpha$-(BEDT-TTF)$_2$I$_3$ samples[3,8,9]. In these studies the authors rely on the X-ray diffraction results under pressure[3] that, using the hopping integrals derived semi-empirically, lead to semimetallic Fermi surfaces via simple tight-binding calculations; the band-crossing Dirac points in this case sink below the Fermi level $E_F$, and a compensating hole pockets emerge in a separate region in the 2D first Brillouin zone. The Dirac points can be adjusted to $E_F$ by introducing either site-dependent potentials or on-site and nearest-neighbour Coulomb repulsions in mean-field theories[3,7], ending up with a pair of charge-neutral tilted Dirac cones as predicted by density functional calculations[8,10] and confirmed by the majority of experiments reported thus far (see Ref. 11 as a review). What if we regard the impact of these Fermi pockets on the behaviour of the NMR relaxation rate? Using the same tight-binding model and the hopping integrals as in Ref. 3, we calculated the expected $T$ dependence of $1/T_1T$ for this semimetallic scenario to check the validity of our analyses solely based on the massless-Dirac-fermion picture.

The 2D tight-binding model is defined as

$$H = \sum_{(i\alpha:j\beta),\sigma} \left( t_{i\alpha:j\beta}\, a^{\dagger}_{i\alpha\sigma} a_{j\beta\sigma} + \text{h.c.} \right), \tag{S2}$$

where $a^{\dagger}_{j\alpha\sigma}$ is the creation operator on the sublattice $j$ = A, A' (= A), B and C in the unit cell $\alpha$ (= $1, \cdots, N_{\text{u.c.}}$) with the spin index $\sigma$ (=↑, ↓), and $t_{i\alpha:j\beta}$ is the nearest-neighbor (NN) hopping energy corresponding to the electron hopping from the $(i, \alpha)$ site to the $(j, \beta)$ site ($N_{\text{u.c.}}$: the total number of unit cells). The value of $t_{i\alpha:j\beta}$ is estimated from the X-ray diffraction data of Ref. 3. The diagonalization of the Hamiltonian yields

$$\sum_{j=1}^{4} \epsilon_{ij}(\mathbf{k})\, d_{j\eta\sigma}(\mathbf{k}) = E_{\eta\sigma}(\mathbf{k})\, d_{i\eta\sigma}(\mathbf{k}), \tag{S3}$$

$$\langle N_{j\sigma}\rangle = \frac{1}{N_{\text{u.c.}}} \sum_{\mathbf{k}} \sum_{\eta=1}^{4} d^*_{j\eta,-\sigma}(\mathbf{k})\, d_{j\eta,-\sigma}(\mathbf{k})\, f\big(E_{\eta,-\sigma}(\mathbf{k}) - \mu\big). \tag{S4}$$

We define $\epsilon_{ij}(\mathbf{k}) = \sum_{\boldsymbol{\delta}_{ij}} t_{ij} e^{i\mathbf{k}\cdot\boldsymbol{\delta}_{ij}}$, where $\boldsymbol{\delta}_{ij}$ is a vector connecting the NN sublattices $i$ and $j$, $E_{\eta\sigma}(\mathbf{k})$ is the eigenvalue ($E_{1\sigma} > E_{2\sigma} > E_{3\sigma} > E_{4\sigma}$), $d_{i\eta\sigma}(\mathbf{k})$ is the corresponding eigenvector with the band index $\eta$, $f(E)$ is the Fermi distribution function and $\mu$ is the chemical potential determined from the condition $\sum_{j\sigma}\langle N_{j\sigma}\rangle = 6$, reflecting the ¾-filling of the electronic band. For simplicity the spin index $\sigma$ is omitted hereafter. For the calculation of the spin susceptibility, we introduce a sublattice spin susceptibility matrix, $\hat{\chi}$, whose $(ij)$-element is given by[1]

$$\chi_{ij}(\mathbf{Q},\omega) = -\frac{1}{N_{\text{u.c.}}} \sum_{\mathbf{k}} \sum_{\eta,\eta'=1}^{4} \mathcal{F}_{ij}^{\eta\eta'}(\mathbf{k},\mathbf{Q}) \frac{f\big(E_\eta(\mathbf{k}+\mathbf{Q})\big) - f\big(E_{\eta'}(\mathbf{k})\big)}{E_\eta(\mathbf{k}+\mathbf{Q}) - E_{\eta'}(\mathbf{k}) - \hbar\omega - i\delta}. \tag{S5}$$

Here, $i\delta$ ($\delta > 0$) is an infinitesimally small imaginary part, and the form factor is defined by

$$\mathcal{F}_{ij}^{\eta\eta'}(\mathbf{k},\mathbf{Q}) = d_{i\eta}(\mathbf{k}+\mathbf{Q}) d^*_{j\eta}(\mathbf{k}+\mathbf{Q}) d_{j\eta'}(\mathbf{k}) d^*_{i\eta'}(\mathbf{k}). \tag{S6}$$

Given that the spin space is isotropic in this model, the longitudinal spin susceptibility ($\hat{\chi}^{\parallel}$) and the transverse susceptibility ($\hat{\chi}^{\perp}$) at the sublattice $i$ are identically given by

$$\chi_i^{\parallel}(\mathbf{Q},\omega) = \chi_i^{\perp}(\mathbf{Q},\omega) = \sum_{j=1}^{4} \chi_{ij}(\mathbf{Q},\omega). \tag{S7}$$

The nuclear spin-lattice relaxation $1/T_1$ at this sublattice is then calculated from Eq. (11) by replacing $\chi_{\perp}(\mathbf{Q},\omega)$ with Eq. (S7).

We insist that the Dirac cones in this semimetallic case are over-tilted such that the gentle slope intersects the Fermi level $E_F$, a situation commonly referred to as the "Type-II" Dirac semimetals, as shown in Supplementary Fig. 2a, b. Remarkably, we find that a saddle-point van Hove singularity appears very close to $E_F$ in the conduction band, locating at the $M(\pi, -\pi)$ point in the first Brillouin zone. Since this singularity leads to a large density of states near $E_F$ (Supplementary Fig. 2c, d), the calculated $1/T_1 T$ [for the sublattice $A$ (= $A'$)] exhibits a monotonic increase towards low $T$ (Supplementary Fig. 2e), in marked contrast to the experimental findings in Fig. 5. This undoubtedly leads us to conclude that the semimetallic scenario is incompatible with our $^{13}$C-NMR data at least in the tested three samples (#1-3), guaranteeing our model analyses assuming only the 2D massless Dirac fermions.

## Supplementary References